\documentclass{eas}
\usepackage{graphicx}
\begin{document}

\title{H$_2$ Distribution during 2-phase Molecular Cloud Formation} \thanks{\emph{Aknowledgements}: CNRS-CONICYT; COSMIS project; PCMI; MesoPSL}

\author{V. Valdivia}
\address{LRA/LERMA, Observatoire de Paris, \'Ecole Normale Sup\'erieure (UMR 8112 CNRS), 24 rue Lhomond, 75231 Paris Cedex 05, France;
\email{valeska.valdivia@obspm.fr}} \secondaddress{Sorbonne Universit\'es, UPMC Univ Paris06, IFD, 4 place Jussieu, 75252 Paris Cedex 05, France}
\author{P. Hennebelle}
\address{Laboratoire AIM, Paris-Saclay, CEA/IRFU/SAp - CNRS - Universit\'e Paris Diderot, 91191 Gif-sur-Yvette Cedex, France} 
\author{M. Gerin}
\sameaddress{1}
\author{P. Lesaffre}
\sameaddress{1}

\begin{abstract}
We performed high-resolution, 3D MHD simulations and we compared to observations of translucent molecular clouds. We show that the observed populations of rotational levels of H$_2$ can arise as a consequence of the multi-phase structure of the ISM.
\end{abstract}
\maketitle

\section{Introduction}
Because the H$_2$ formation time is long (some $10^9/n ~\mathrm{yrs}$, where $n$ is the density in $\mathrm{cm^{-3}}$), and can exceed the crossing time of molecular clouds ($t_\mathrm{cross} = 10^6 (\frac{r}{1~\mathrm{pc}})^{0.5}~\mathrm{yrs}$), it is important to consider the impact of dynamical effects on the evolution of the H$_2$ fraction.  To understand the H$_2$ molecule formation process under the dynamical influence of a highly inhomogeneous structure, such as that of molecular clouds, where  warm and cold phases are interwoven, we performed high-resolution MHD simulations of realistic molecular clouds formed through colliding streams of warm atomic gas. For this study we used the RAMSES code (Teyssier \cite{teyssier2002}), where we included the formation and destruction processes for H$_2$, as well as the thermal feedback (see Valdivia \etal\ \cite{valdivia2015}). The effects of dust shielding for the UV radiation, and the self-shielding due to H$_2$ molecules were included by using our tree-based method, detailed in Valdivia \& Hennebelle (\cite{valdivia2014}).
\section{Results}
\begin{figure}[h]
\includegraphics[height=6.7cm]{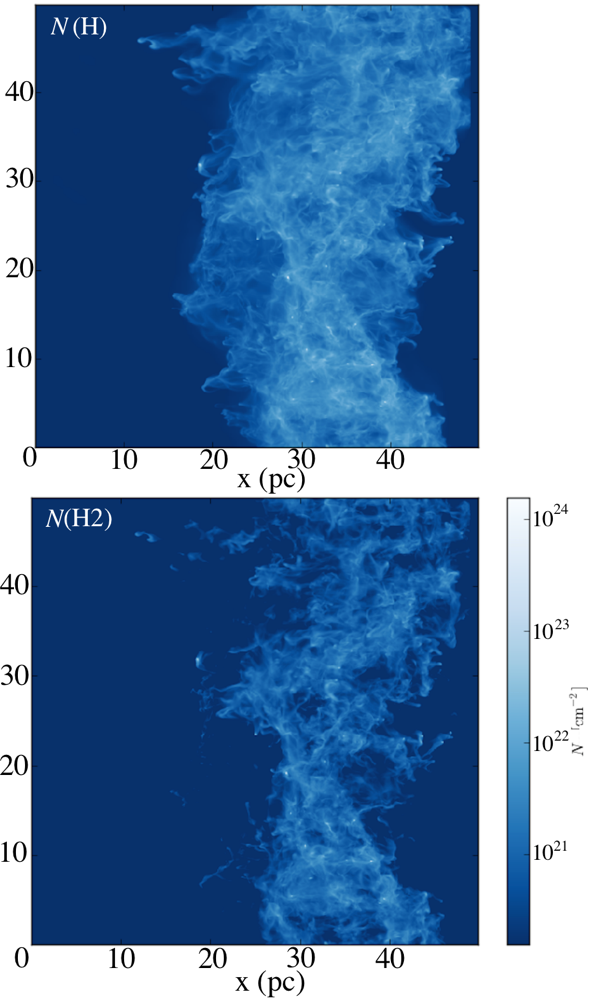}
\hfill
\includegraphics[height=6.8cm]{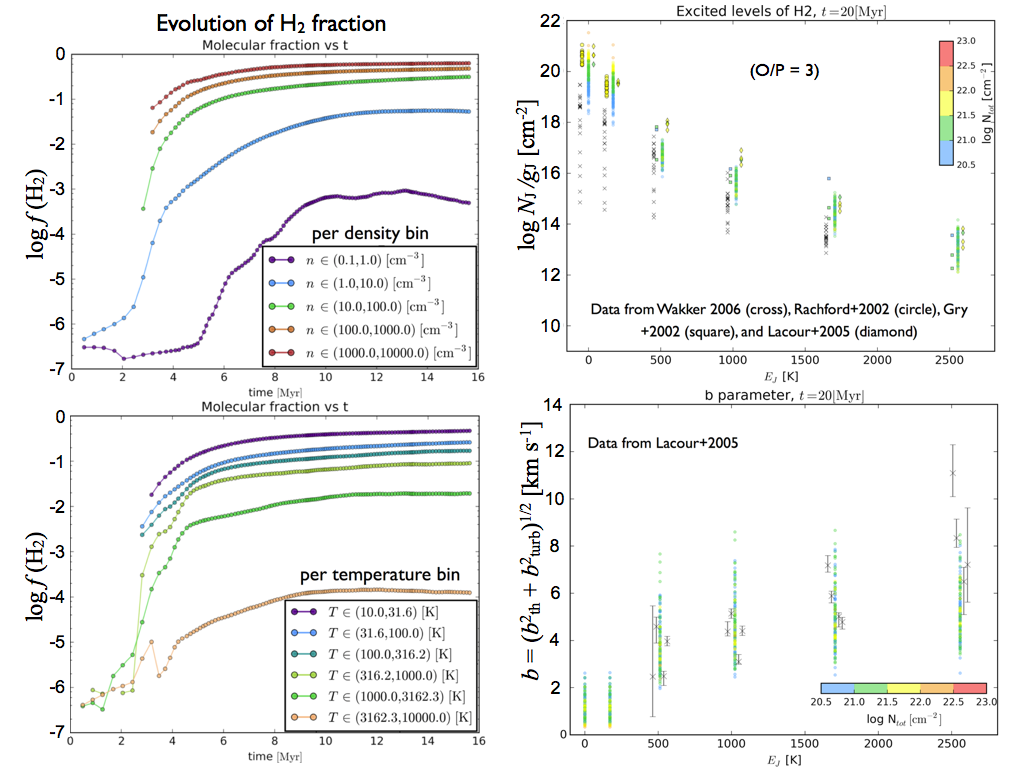}
\caption{\emph{Left}: Total H, and H$_2$ column densities. \emph{Center}: H$_2$ fraction evolution per density bin and per temperature bin. \emph{Right}: H$_2$ population distribution, and $b$ parameter.}
\label{fig1}
\end{figure}

H$_2$ is mainly formed in dense regions (Fig. \ref{fig1} \emph{left}), and  it is formed faster than the usual estimates based on the mean density of the cloud  (Fig. \ref{fig1} \emph{center}), about $4~\mathrm{Myr}$ instead of $20~\mathrm{Myr}$ for $\bar n \sim 50~\mathrm{cm^{-3}}$. We interpret this as the result of the combined effect of the local density enhancement, that drives a faster H$_2$ formation (see also Glover \& MacLow \cite{glover2007}), and the shielding provided by the global structure, that ensures the survival of molecular gas. Additionally, the mixing between different phases induce the presence of H$_2$ molecules in the warm phase, explaining the observed warm H$_2$ in the diffuse ISM.
We have calculated the H$_2$ populations in the first levels at thermal equilibrium, as well as the Doppler broadening parameter $b$ for each population (Fig. \ref{fig1} \emph{right}). Excitation diagrams and $b$ parameters are in good agreement with the values observed by Copernicus and \emph{FUSE} (Lacour \cite{lacour2005}, Gry \etal\ \cite{gry2002}, Rachford \etal\ \cite{rachford2002}, Wakker \cite{wakker2006}). Altogether, these results suggest that excited populations might be the consequence of local temperature and density within molecular clouds, highlighting the multi-phase nature of molecular clouds.

\end{document}